\documentclass[twocolumn,prl,aps,showpacs,preprintnumbers,amsmath,amssymb, superscriptaddress]{revtex4-1}
\pdfoutput=1

\usepackage{bm}
\usepackage{graphicx}
\usepackage{color}

\begin{document}

\title{Possible nodal superconducting gap emerging at the Lifshitz transition in heavily hole-doped Ba$_{0.1}$K$_{0.9}$Fe$_2$As$_2$}

\author{N. Xu}\email{nan.xu@psi.ch}
\affiliation{Beijing National Laboratory for Condensed Matter Physics, and Institute of Physics, Chinese Academy of Sciences, Beijing 100190, China}
\affiliation{Paul Scherrer Institut, Swiss Light Source, CH-5232 Villigen PSI, Switzerland}
\author{P. Richard}\email{p.richard@iphy.ac.cn}
\affiliation{Beijing National Laboratory for Condensed Matter Physics, and Institute of Physics, Chinese Academy of Sciences, Beijing 100190, China}
\author{X. Shi}
\affiliation{Beijing National Laboratory for Condensed Matter Physics, and Institute of Physics, Chinese Academy of Sciences, Beijing 100190, China}
\affiliation{Paul Scherrer Institut, Swiss Light Source, CH-5232 Villigen PSI, Switzerland}
\author{A. van Roekeghem}
\affiliation{Beijing National Laboratory for Condensed Matter Physics, and Institute of Physics, Chinese Academy of Sciences, Beijing 100190, China}
\affiliation{Centre de Physique Th\'{e}Žorique, Ecole Polytechnique, CNRS-UMR7644, 91128 Palaiseau, France}
\author{T. Qian}
\affiliation{Beijing National Laboratory for Condensed Matter Physics, and Institute of Physics, Chinese Academy of Sciences, Beijing 100190, China}
\author{E. Razzoli} 
\affiliation{Paul Scherrer Institut, Swiss Light Source, CH-5232 Villigen PSI, Switzerland}
\author{E. Rienks} 
\affiliation{Helmholtz-Zentrum Berlin, BESSY, D-12489 Berlin, Germany}
\author{G.-F. Chen}
\affiliation{Beijing National Laboratory for Condensed Matter Physics, and Institute of Physics, Chinese Academy of Sciences, Beijing 100190, China}
\author{E. Ieki}
\affiliation{Department of Physics, Tohoku University, Sendai 980-8578, Japan}
\author{K. Nakayama}
\affiliation{Department of Physics, Tohoku University, Sendai 980-8578, Japan}
\author{T. Sato}
\affiliation{Department of Physics, Tohoku University, Sendai 980-8578, Japan}
\author{T. Takahashi}
\affiliation{Department of Physics, Tohoku University, Sendai 980-8578, Japan}
\affiliation{WPI Research Center, Advanced Institute for Materials Research, Tohoku University, Sendai 980-8577, Japan}
\author{M. Shi}
\affiliation{Paul Scherrer Institut, Swiss Light Source, CH-5232 Villigen PSI, Switzerland}
\author{H. Ding}\email{dingh@iphy.ac.cn}
\affiliation{Beijing National Laboratory for Condensed Matter Physics, and Institute of Physics, Chinese Academy of Sciences, Beijing 100190, China}

\date{\today}

\begin{abstract}
We performed a high energy resolution ARPES investigation of over-doped Ba$_{0.1}$K$_{0.9}$Fe$_2$As$_2$ with $T_c= 9$ K. The Fermi surface topology of this material is similar to that of KFe$_2$As$_2$ and differs from that of slightly less doped Ba$_{0.3}$K$_{0.7}$Fe$_2$As$_2$, implying that a Lifshitz transition occurred between $x=0.7$ and $x=0.9$. Albeit for a vertical node found at the tip of the emerging off-M-centered Fermi surface pocket lobes, the superconducting gap structure is similar to that of Ba$_{0.3}$K$_{0.7}$Fe$_2$As$_2$, suggesting that the paring interaction is not driven by the Fermi surface topology. 
\end{abstract}


\maketitle

The discovery of high-temperature superconductivity without hole Fermi surface (FS) pocket \cite{JG_Guo_PRB2010,Qian_PRL2011,Y_Zhang_NatureMat2011, D_MouPRL2011, XP_WangEPL2011, ZH_LiuPRL2012, XP_WangEPL2012} in AFe$_2$Se$_2$ (A = Tl, K, Cs, Rb) imposed severe constraints to the electron-hole quasi-nesting scenario as the main Cooper pairing force in the Fe-based superconducting (SC) materials \cite{Richard_RoPP2011} and raised fundamental questions related to the importance of their FS topology. To answer these questions, it is necessary to investigate heavily hole-doped compounds. Previous angle-resolved photoemission spectroscopy (ARPES) studies of fully hole-doped KFe$_2$As$_2$ indicate that the FS near the M$(\pi,0)$ point is formed by small off-M-centered hole FS pocket lobes \cite{Sato_PRL2009,Yoshida_JCPS72} rather than the M-centered ellipsoid-like electron FS pockets commonly observed in the other materials \cite{Richard_RoPP2011}. Interestingly, KFe$_2$As$_2$ has a very low critical temperature ($T_c$) of only 3 K \cite{Rotter_Angew2008,H_Chen_EPL85}, which was early \cite{Sato_PRL2009} interpreted as, and is still widely considered as, a consequence of the evolution of the FS topology. Despite their incapability to access the band structure at the M point and thus to reveal completely the SC gap structure, laser-ARPES measurements suggest a rather complicated nodal SC gap profile around the $\Gamma$ point \cite{Shin_KFA_Science}, which is consistent with the finite residual thermal conductivity ($\kappa_0/T$) of this material at low temperature \cite{JK_DongPRL2010,ReidPRL109}.

While their existence is widely accepted, the origin of the nodes in KFe$_2$As$_2$ and their relationship with the pairing mechanism remain unclear and could possibly involve a fundamental change in the SC order parameter upon doping K from the optimal Ba$_{0.6}$K$_{0.4}$Fe$_2$As$_2$ composition, for which both the ARPES \cite{Ding_EPL83,L_Zhao_CPL2008,Nakayama_EPL2009} and thermal conductivity \cite{XG_LuoPRB2009} techniques agree on a nodeless SC gap. Indeed, Tafti \emph{et al.} \cite{Tafti_NPhys2013} recently reported a sudden reversal in the pressure ($P$) dependence of $T_c$ in KFe$_2$As$_2$ without discontinuity in the Hall coefficient $R_H(P)$ and in the electrical resistivity $\rho(P)$, which was interpreted as an evidence for a change in the order parameter incompatible with a Lifshitz transition (change in the FS topology \cite{Lifshitz_JETP11}). Based on a rigid-band shift model, the Lifshitz transition corresponding to the apparition of the small off-M-centered hole FS pocket lobes was estimated to occur within the $0.8\leq x\leq 0.9$ doping range \cite{Nakayama_PRB2011}. Determining the $k$-space structure of the SC gap in the vicinity of this transition is of critical importance.

In this Letter, we investigate the electronic structure and SC gap of heavily hole-doped Ba$_{0.1}$K$_{0.9}$Fe$_2$As$_2$ in the vicinity of the Lifshitz transition associated with the FS topology around the M point. While the FS topology at the $\Gamma$ point remains unchanged as compared that of Ba$_{0.3}$K$_{0.7}$Fe$_2$As$_2$ \cite{Nakayama_PRB2011}, four small off-M-centered hole FS pocket lobes emerge in Ba$_{0.1}$K$_{0.9}$Fe$_2$As$_2$. Despite such strong modification of the FS topology, the $T_c$ of this material remains as high as 9 K around this particular doping for which no big jump of $T_c$ has been reported \cite{Rotter_Angew2008,H_Chen_EPL85}. As with Ba$_{0.3}$K$_{0.7}$Fe$_2$As$_2$, we observe rather isotropic SC gaps on each hole FS around the zone center. However, a vertical node is observed at the tip of the emerging off-M-centered FS pocket lobes. Our results suggest that the paring interaction is not driven by the Fermi surface topology.

Large single crystals of Ba$_{0.1}$K$_{0.9}$Fe$_2$As$_2$ were grown by the self-flux method \cite{GF_ChenPRB78}. ARPES measurements were performed at Swiss Light Source surface/interface spectroscopy (SIS) beamline and at the 1-cubed ARPES end-station of BESSY using VG-Scienta R4000 electron analyzers with photon energy ranging from 20 to 80 eV. The angular resolution was set to 0.2$^{\circ}$ and the energy resolution to 2$\sim$10 meV for SC gap and band structure measurements, respectively. Clean surfaces for the ARPES measurements were obtained by cleaving crystals \emph{in situ} in a working vacuum better than 5 $\times$ 10$^{-11}$ Torr. We label the momentum ($k$) values with respect to the 1 Fe/unit cell Brillouin zone (BZ), and use $c'=c/2$ as the distance between two Fe planes.


\begin{figure}[!t]
\begin{center}
\includegraphics[width=3.4in]{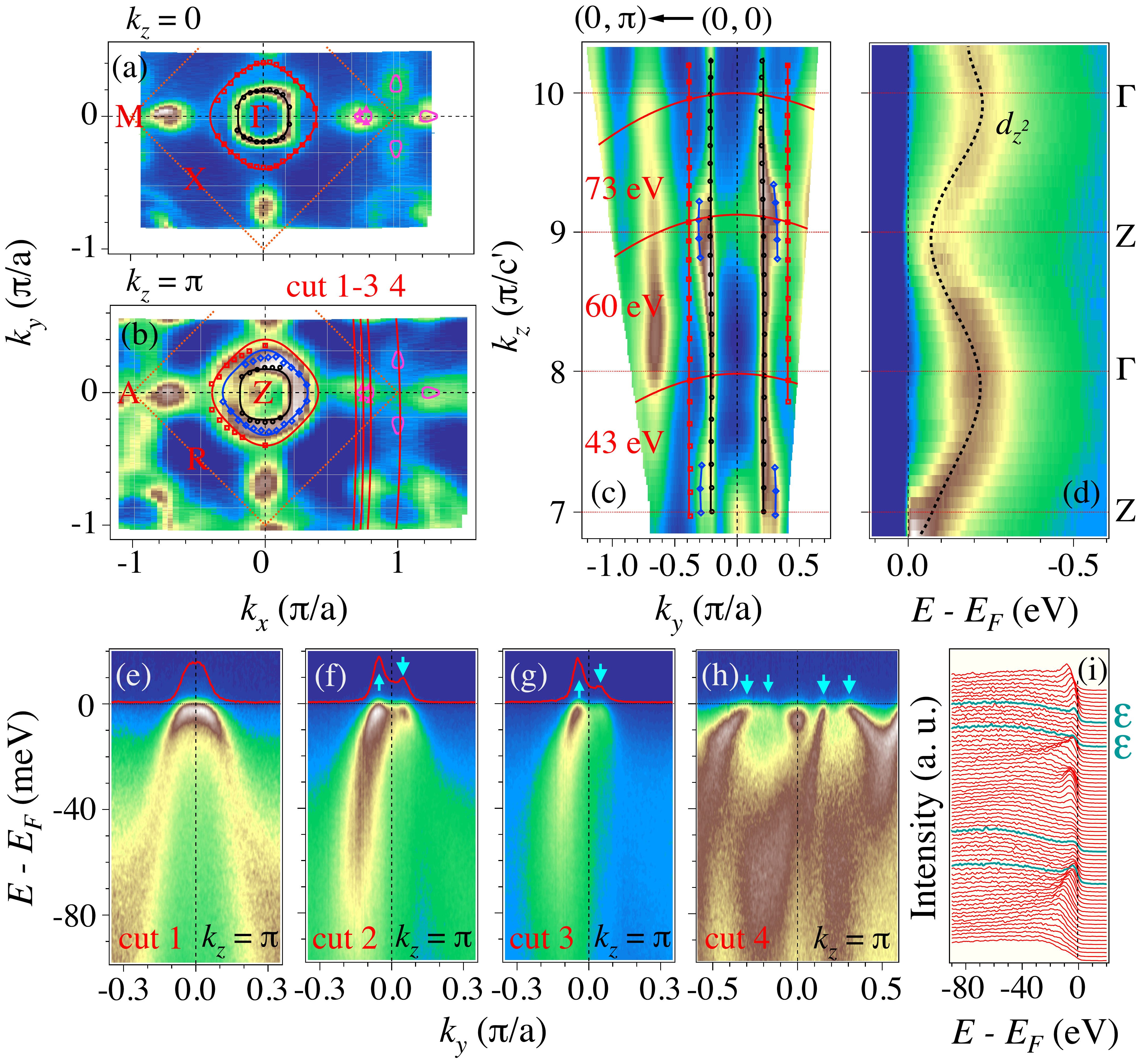}
\end{center}
\caption{\label{Fig1_FS}(Color online): (a) ARPES FS intensity mapping ($\pm 5$ meV integration) recorded with $h\nu = 43$ eV ($k_z = 0$). (b) Same as (a) but with $h\nu = 60$ eV ($k_z = \pi$). Symbols indicate the $k_F$ positions the labeled red lines in (b) indicate ARPES cuts shown in (e)-(h). (c) ARPES FS intensity mapping in the $k_x-k_z$ plane. (d) ARPES intensity cut along $\Gamma$-Z. (e)-(g) ARPES intensity plot recorded with $h\nu=31$ eV ($k_z = \pi$) along the $\varepsilon$ pockets (cut 1-3 in (b)). The red curve are the MDCs at $E_F$. (h) ARPES intensity plot recorded with $h\nu=31$ eV ($k_z = \pi$) along the Z-A direction, in the vicinity of the A point (cut 4 in (b)). (i) Corresponding EDCs. }
\end{figure}

We show in Figs. \ref{Fig1_FS}(a) and \ref{Fig1_FS}(b) the ARPES FS intensity mappings recorded with $h\nu = 43$ eV and $h\nu = 60$ eV, respectively. As determined from our $h\nu$ analysis, these $h\nu$ values correspond respectively to out-of-plane momentum ($k_z$) values equivalent to 0 and $\pi$. Similarly to optimally-doped Ba$_{0.6}$K$_{0.4}$Fe$_2$As$_2$ \cite{Ding_EPL83,L_Zhao_CPL2008} and over-doped Ba$_{0.3}$K$_{0.7}$Fe$_2$As$_2$ \cite{Nakayama_PRB2011}, three hole FS pockets are distinguished at the BZ center ($\Gamma$), two of them being nearly degenerate at $k_z = 0$. Instead of M-centered ellipsoid electron FS pockets, the intensity mappings exhibit off-M-centered lobes slightly elongated along the $\Gamma$-M axis, which correspond to the $\varepsilon$ FS pockets in Ref. \cite{Sato_PRL2009}. Below we show that in contrast to over-doped Ba$_{0.3}$K$_{0.7}$Fe$_2$As$_2$ \cite{Nakayama_PRB2011} and similarly to KFe$_2$As$_2$ \cite{Sato_PRL2009,Yoshida_JCPS72}, this intensity comes from a band crossing the Fermi level ($E_F$) rather than a band topping slightly below $E_F$. We caution that some weak intensity is also found at the M point. Supported by our band structure calculations illustrated in Fig. \ref{Fig4_model}, we attribute this intensity to the tail of the electron bands pushed above $E_F$ following hole-doping. However, we cannot completely rule out the possibility that one electron band is still crossing $E_F$ and forming a tiny electron FS pocket at M. 

The dispersion obtained by converting $h\nu$ into $k_z$ using the nearly-free electron approximation \cite{DamascelliPScrypta2004} with an inner potential of 12.5 eV is displayed in Fig. \ref{Fig1_FS}(c). It shows small $k_z$ warping of the FS, especially for the quasi-degenerate inner hole bands centered at $\Gamma$. This $k_z$ warping is small for the $\varepsilon$ band, confirming that the FS topology does not depend on the probing $h\nu$ value. Using the full set of ARPES data and the Luttinger theorem, we estimate the hole concentration to be $\sim 0.42$ per Fe, which is close to the nominal composition. Interestingly, the  $k_x-k_z$ mapping displayed in Fig. \ref{Fig1_FS}(c) shows some additional ARPES intensity at the Z$(0,0,\pi)$ point. This feature originates from a highly $k_z$-dispersive hole band, most likely of $d_{z^2}$ orbital character, whose top oscillates from -40 meV to -220 meV, as illustrated in Fig. \ref{Fig1_FS}(d).   

We present a more detailed analysis of the $\varepsilon$ pockets in Figs. \ref{Fig1_FS} (e)-(g), using ARPES cuts perpendicular to the Z-A direction for the $k_z$ = $\pi$ plane (cuts 1-3 from Fig. \ref{Fig1_FS}(b)). While the band is not crossing $E_F$ along cut 1, as confirmed by the momentum distribution curve (MDC) at $E_F$, we identify $k_F$ vectors separated by 0.05 $\pi/a$ along cut 2, indicating that the top of the $\varepsilon$ band has emerged. The spacing becomes slightly smaller, 0.04 $\pi/a$, along cut 3. The ARPES intensity cut along Z-A for the $k_z = \pi$ plane (cuts 4 from Fig. \ref{Fig1_FS}(b)) is shown in Fig. \ref{Fig1_FS} (h), and the corresponding energy dispersive curves (EDCs) are shown in Fig. \ref{Fig1_FS} (i). Clearly, the $\beta$ band turns up around $k_y = \pm 0.50$ $\pi/a$ and crosses $E_F$ to form the small hole $\varepsilon$ FS with the $\gamma$ band, as in KFe$_2$As$_2$ \cite{Sato_PRL2009,Yoshida_JCPS72}. As mentioned above,  also observe some intensity at the A point attributed to the band bottom of the electron-like $\delta$ band slightly above $E_F$.

\begin{figure}[!t]
\begin{center}
\includegraphics[width=3.4in]{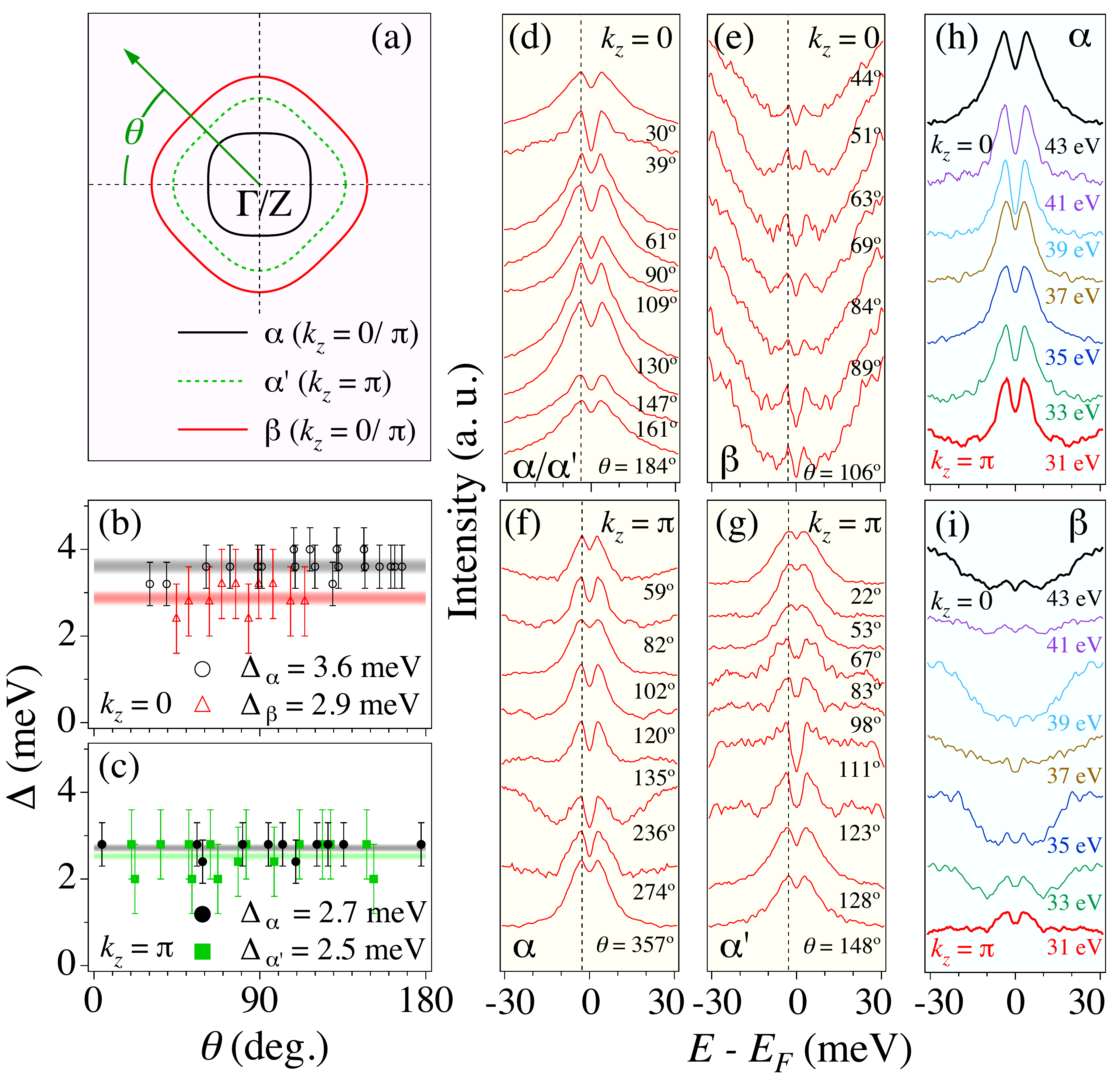}
\end{center}
\caption{\label{Fig2_SCgap_G}(Color online) (a) Schematic FS around the BZ center and definition of the polar angle $\theta$. The axes coincide with the $\Gamma$-M directions. (b) SC gap size as a function of $\theta$ at 0.9 K for the nearly degenerate $\alpha$/$\alpha$' FSs and the $\beta$ FS, in the $k_z = 0$ plane. (c) Same as (b), but for the $\alpha$ and $\alpha$' FSs for the $k_z$ = $\pi$ plane. (d) Symmetrized EDCs for different polar angles of the $\alpha$/$\alpha$' bands in the $k_z = 0$ plane. (e) Same as (d), but for the $\beta$ band. (f)-(g) Symmetrized EDCs for different polar angles of the $\alpha$ and $\alpha$' bands in the $k_z= \pi$ plane, respectively. (f)-(g) EDC plots in the $k_z= \pi$ plane for the $\alpha$, $\alpha$' and $\beta$ bands, respectively. (h)-(i) Symmetrized EDCs at different $h\nu$ values for the $\alpha$ and $\beta$ bands, respectively.
}
\end{figure}

\begin{figure}[!t]
\begin{center}
\includegraphics[width=3.4in]{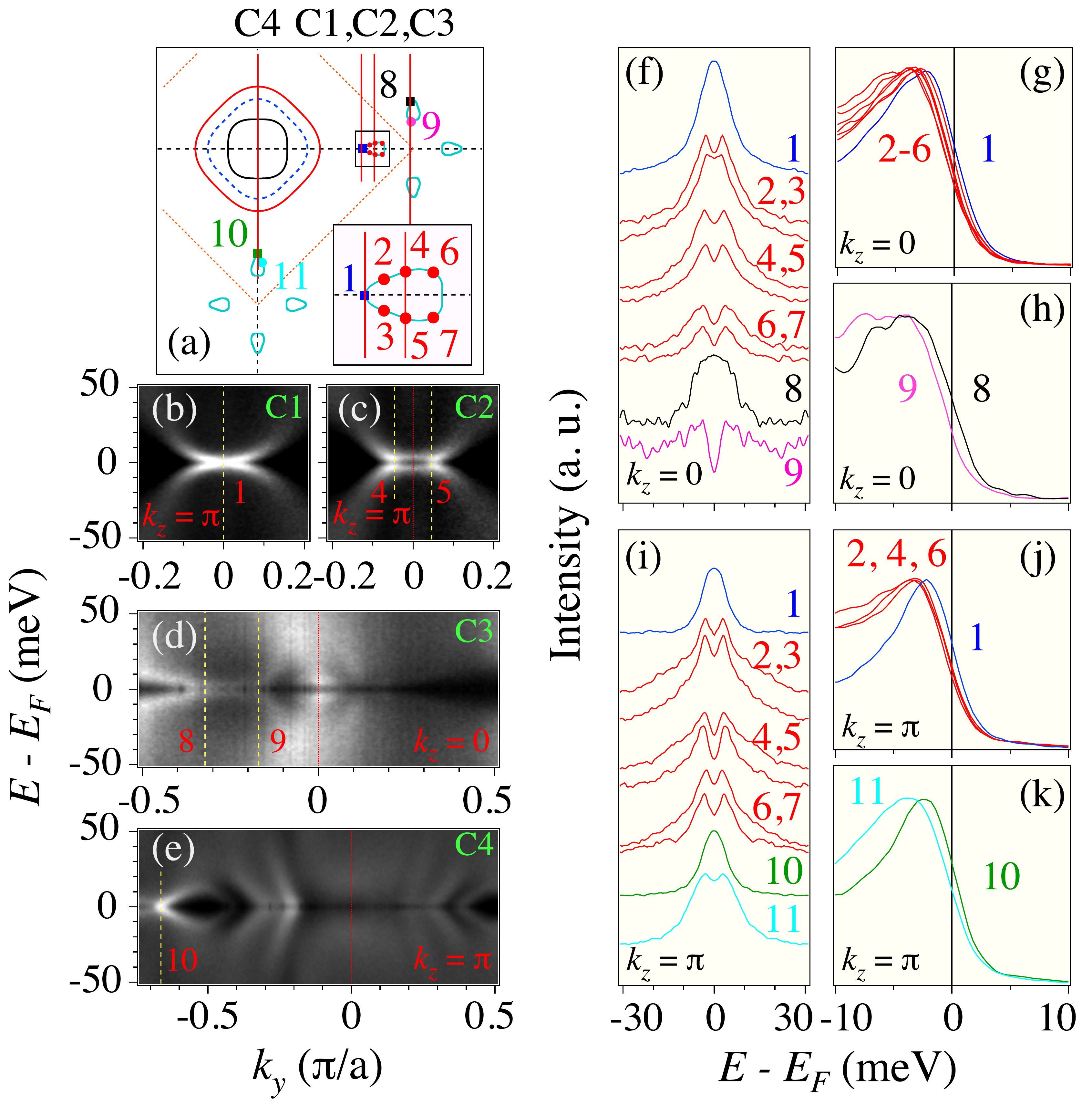}
\end{center}
\caption{\label{Fig3_SCGap_eta}(Color online) (a) Momentum location of ARPES cuts (labeled red lines) shown in (b)-(e). The symbols indicate $k_F$ positions of the EDCs and symmetrized EDCs in (f)-(k). (b)-(e) Symmetrized ARPES intensity plots for the C1, C2, C3 and C4 cuts. The labeled yellow lines are $k_F$ positions of the EDCs and symmetrized EDCs in (f)-(k). (f) Symmetrized EDCs along the $\varepsilon$ FS at $k_z = 0$. (g)-(h) EDCs along the $\varepsilon$ FS at $k_z = 0$. (i)-(k) Same as (f)-(h) but for $k_z = \pi$.
}
\end{figure}

As predicted from our previous data \cite{Nakayama_PRB2011}, our Ba$_{0.1}$K$_{0.9}$Fe$_2$As$_2$ results clearly show that a Lifshitz transition occurs between $x=0.7$ and $x=0.9$ upon hole-doping the Ba$_{1-x}$K$_{x}$Fe$_2$As$_2$ system. In order to investigate the interplay between the FS topology and the SC gap structure, we conducted high-energy resolution ARPES measurements at $T$ = 0.9 K, way below the SC transition. In Fig. \ref{Fig2_SCgap_G}, we plot the SC gap results for FSs located at the BZ center ($\Gamma$/Z). The summaries of the SC gap amplitudes as a function of the polar angle $\theta$ defined in Fig. \ref{Fig2_SCgap_G}(a) are shown in Fig. \ref{Fig2_SCgap_G}(b) and Fig. \ref{Fig2_SCgap_G}(c) for the $k_z=0$ and $k_z=\pi$ planes, respectively. The corresponding symmetrized EDCs are displayed in Figs. \ref{Fig2_SCgap_G}(d)-\ref{Fig2_SCgap_G}(g). Within experimental uncertainties, a constant SC gap of $\Delta^{k_z = 0}_{\alpha/\alpha'} = 3.6$ meV is extracted along the $\alpha/\alpha'$ FSs in the $k_z = 0$ plane. A slightly smaller but still isotropic gap of $\Delta^{k_z = 0}_{\beta} = 2.7$ meV is extracted in the same plane for the $\beta$ band. In the $k_z=\pi$ plane, for which the $\alpha$ and $\alpha$' bands can be resolved, our results indicate isotropic SC gaps with almost identical sizes of $\Delta^{k_z = \pi}_{\alpha} = 2.7$ meV and $\Delta^{k_z = \pi}_{\alpha'} = 2.5$ meV, respectively. Unfortunately, the weakness of the coherence peak associated with the $\beta$ band near $k_z = \pi$ does not allow a reliable determination of the SC gap. Nevertheless, our $h\nu$ dependent measurements along $\Gamma$-M suggest that the SC gap size on all the $\Gamma$-centered hole FSs does not vary too much along $k_z$.

\begin{figure}[!t]
\begin{center}
\includegraphics[width=3.4in]{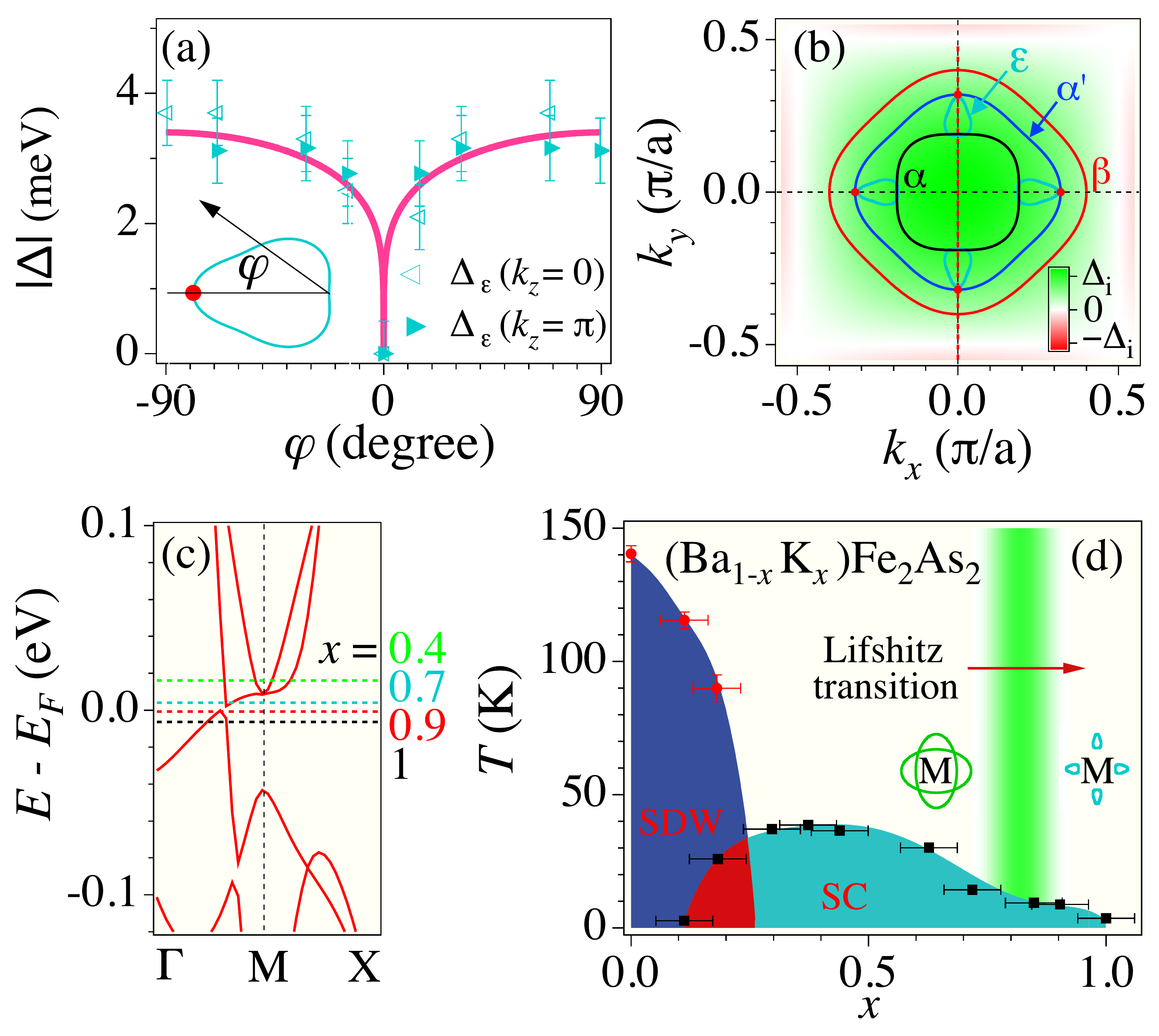}
\end{center}
\caption{\label{Fig4_model}(Color online) (a) SC gap size  at 0.9 K along the $\varepsilon$ FS as a function of the angle $\varphi$ (define in inset). The pink line is a guide for the eye. (b) FS of Ba$_{0.1}$K$_{0.9}$Fe$_2$As$_2$ with the $\varepsilon$ FS pockets shifted by $(-\pi, 0)$. The color scale represents the amplitude of the $\cos k_x \cos k_y$ global pairing function. (c) LDA band structure calculations from Ref. \cite{G_Xu_EPL2008}, renormalized by a factor 2. The location of the chemical potential is indicated for several doping levels. (d) Location of the Lifshitz transition in Ba$_{1-x}$K$_x$Fe$_2$As$_2$ with respect to the phase diagram extracted from Ref. \cite{Rotter_Angew2008}.}
\end{figure}

Now we focus on the SC gap of the $\varepsilon$ pockets, for which the $k$ location of the measured cuts and points are shown in Fig. \ref{Fig3_SCGap_eta}(a). In Figs. \ref{Fig3_SCGap_eta} (b)-(e), we display the symmetrized ARPES intensity at $T = 0.9$ K for various cuts in the $k_z = \pi$ plane, and the symmetrized EDCs along the $\varepsilon$ FS at several $k_F$ positions indicated in Fig. \ref{Fig3_SCGap_eta}(a) are plotted in Figs. \ref{Fig3_SCGap_eta}(f) and \ref{Fig3_SCGap_eta}(i) for the $k_z$ = 0 and $\pi$ planes, respectively. Interestingly, a SC gap opens along the $\varepsilon$ pockets, except at points labeled 1, 8 and 10, which correspond to the tip position of the $\varepsilon$ lobes. As shown in Figs. \ref{Fig3_SCGap_eta}(g)-\ref{Fig3_SCGap_eta}(h) and Figs. \ref{Fig3_SCGap_eta}(j)-\ref{Fig3_SCGap_eta}(k) for $k_z=0$ and $k_z=\pi$ planes, respectively, these SC nodes are confirmed by the absence of leading edge gap in the related EDCs as compared to the other locations.

The angular distribution of the SC gap along the $\varepsilon$ FS is shown in Fig. \ref{Fig4_model}(a). To check the origin of the SC gap anomaly at the tip position ($\varphi=0$), we display in Fig. \ref{Fig4_model}(b) the FS of this material in the $k_z=\pi$ plane after shifting the $\varepsilon$ FSs lobes by ($-\pi$, 0). For convenience, we added a color scale proportional to the amplitude of the $|\cos(k_x)\cos(k_y)|$ global pairing function expected from the strong coupling approach and the effective $J_1-J_2$ model \cite{SeoPRL2008,C_Fang_PRX2011,HuJP_SR2012,ZhouYi_EPL2011,HuJP_PRX2012}. While the absence of node around the $\Gamma$ point is inconsistent with a $d$-wave gap symmetry and matches well the global gap function, our finding of nodes on the $\varepsilon$ FS cannot be explained by this simple gap function since the $\varepsilon$ FS pockets locate far from the nodal lines. Interestingly, the superimposition of the $\varepsilon$ FS with the $\Gamma$-centered FSs may offer possible hints to explain the origin of the nodes. Indeed, while the flatter part of the $\varepsilon$ FS near $\varphi=90$ degrees is connected to the $\alpha$ FS by the ($\pi$, 0) wave vector, the tip of the ellipsoid is linked to the $\alpha$' FS by the same wave vector. Although the details of the low-energy scattering would depend on the orbital components at each FS point, similar effect has been invoked to explain the departure from the $|\cos(k_x)\cos(k_y)|$ gap function in LiFeAs \cite{UmezawaPRL108}. 

A Lifshitz transition has been evidenced at low electron-doping in the Ba(Fe$_{1-x}$Co$_x$)$_2$As$_2$ system. Due to long-range antiferromagnetic ordering, the non-SC parent compound BaFe$_2$As$_2$ undergoes a non-trivial reconstruction of its electronic band structure characterized by Dirac cones \cite{RanPRB79, RichardPRL2010,Harrison_PRB80}. Upon electron-doping, a previous study indicates that the emergence of superconductivity occurs when this band structure reconstruction disappears \cite{C_LiuNaturePhys2010}. Our results prove that a Lifshitz transition also takes place upon hole-doping the Ba$_{1-x}$K$_{x}$Fe$_2$As$_2$ system from $x=0.7$ to $x=0.9$. Within the framework of the electron-hole quasi-nesting scenario \cite{MazinPhysicaC2009, Graser_NJP2009,Richard_RoPP2011}, the replacement of the M-centered electron FS pockets by the hole $\varepsilon$ FS pockets should be armful to superconductivity. Unless the electron FS pockets do not play any important role in the Cooper pairing, we argue that the Lifshitz transition should also have a large impact on the SC properties determined from other weak coupling scenarios, for example by changing significantly the intra-pocket and inter-pocket scattering involved in some models \cite{MaitiPRB85}. In contrast, pairing described by strong coupling approaches \cite{SeoPRL2008, C_Fang_PRX2011, HuJP_SR2012, ZhouYi_EPL2011, HuJP_PRX2012} should suffer smaller influence. 

To put our findings in context, we digitized the early phase diagram of the Ba$_{1-x}$K$_x$Fe$_2$As$_2$ system reported by Rotter \emph{et al.} \cite{Rotter_Angew2008} in Fig. \ref{Fig4_model}(d). Albeit for a small slope change around $x=0.8$, the critical temperature $T_c(x)$ evolves smoothly on the overdoped side of the phase diagram. According to our Local Density Approximation (LDA) calculations reproduced in Fig. \ref{Fig4_model}(c), this slope change can be attributed to the Lifshitz transition associated with the emergence of the $\varepsilon$ pockets. While the FS topology is changed, $T_c$ remains relatively high, which contrasts with intuitive description of the SC properties from a weak-coupling scenario and suggests that the driving SC force is the same throughout the whole phase diagram. Moreover, it suggests that the Lifshitz transition observed here has fundamentally different nature from the one observed in Ba(Fe$_{1-x}$Co$_x$)$_2$As$_2$ \cite{C_LiuNaturePhys2010}, where long-range antiferromagnetism is suppressed.

Our findings also allow us to interpret the recent discovery of a sign reversal in the $P$ dependence of $T_c$ in KFe$_2$As$_2$ with perspectives different from that of Tafti \emph{et al.} \cite{Tafti_NPhys2013}. In their analysis, Tafti \emph{et al.} adopt a weak-coupling approach and neglect the existence of the $\varepsilon$ FS lobes, thus neglecting the possibility of having nodes on that particular FS. Assuming that this node is induced by low-energy interband scattering, it is plausible that the FS evolves only slightly, without any Lifshitz transition or large size variation of the FS pockets, so that the interband scattering between the tip of the $\varepsilon$ lobes and the $\alpha$' band is suppressed, leading to a sign reversal in the $P$ dependence of $T_c$ without jump in $R_H(P)$ or $\rho(P)$. Alternatively, the value of $T_c$ within the strong coupling approach is controlled by local $J$ parameters, which are themselves dependent on the Fe-As bounding. In this context, it is natural to encounter extrema in the Fe-As bounding configuration as a function of $P$, thus explaining the sign reversal in the $P$ dependence of $T_c$ without invoking modification of the FS topology or even a change in the symmetry of the order parameter. Yet, further studies are necessary to confirm or infirm the original interpretation of Tafti  \emph{et al.}.

In summary, we showed that the FS topology of over-doped Ba$_{0.1}$K$_{0.9}$Fe$_2$As$_2$ with $T_c=9$ K is similar to that of KFe$_2$As$_2$. Since $T_c$ does not change significantly across the corresponding Lifshitz transition, we infer from our results that the pairing mechanism is unlikely driven by the FS topology, at least around the M point. We observe a SC gap opening at $T = 0.9$ K, with nodes at the tips of the emerging $\varepsilon$ pockets. Our result suggests that the SC gap symmetry is still $S_{\pm}$, instead of $d$-wave, and is more consistent with a strong coupling approach.

We thank F. F. Tafti, L. Taillefer, X. Dai, J.-P. Hu and N. L. Wang for useful discussions. This work was supported by grants from CAS (2010Y1JB6), MOST (2010CB923000 and 2011CBA001000, 2011CBA00102, 2012CB821403) and NSFC (10974175, 11004232, 11050110422 and 11034011/A0402) of China and from JSPS and MEXT from Japan. This work was also supported by the Sino-Swiss Science and Technology Cooperation (project no. IZLCZ2 138954), the Swiss National Science Foundation (No. 200021-137783). This work is based in part upon research conducted at the Swiss Light Source, Paul Scherrer Institut, Villigen, Switzerland.

\bibliography{biblio_ens}

\end{document}